\documentclass{svjour3}                     
\smartqed  
\usepackage{graphicx}
\begin{document}

\title{Form Factors of Few-Body Systems: \thanks{Presented by Mar\'ia G\'omez-Rocha at LIGHTCONE 2011, Dallas, USA, 23 - 27 May, 2011.}
}
\subtitle{Point Form Versus Front Form}


\author{M. G\'omez-Rocha, E. P. Biernat, W. Schweiger}


\institute{M. G\'omez Rocha \and W. Schweiger \at
              Institut f\"ur Physik, FB Theoretische Physik, Universit\"at Graz, A-8010 Graz, Austria \\
\email{maria.gomez-rocha@uni-graz.at} \\
\email{wolfgang.schweiger@uni-graz.at}\\
E. P. Biernat \at
             Physics Department, IST, Universidade T\'ecnica de Lisboa,
1049-001 Lisboa, Portugal \\
              \email{elmar.biernat@ist.utl.pt}}

\date{Received: date / Accepted: date}

\maketitle

\begin{abstract}
We present a relativistic point-form approach for the calculation
of electroweak form factors of few-body bound states that leads to
results which resemble those obtained within the covariant
light-front formalism of Carbonell et al. Our starting points are
the physical processes in which such form factors are measured,
i.e. electron scattering off the bound state, or the semileptonic
weak decay of the bound state. These processes are treated by
means of a coupled-channel framework for a Bakamjian-Thomas type
mass operator. A current with the correct covariance properties is
then derived from the pertinent leading-order electroweak
scattering or decay amplitude. As it turns out, the
electromagnetic current is affected by unphysical contributions
which can be traced back to wrong cluster properties inherent in
the Bakamjian-Thomas construction. These spurious contributions,
however, can be separated uniquely, as in the covariant
light-front approach. In this way we end up with form factors
which agree with those obtained from the covariant
light-front approach. As an example we will present results for
electroweak form factors of heavy-light systems and discuss the
heavy-quark limit which leads to the famous Isgur-Wise function.
\end{abstract}

\section{Introduction}
\label{intro} Electroweak processes in which either an electron is
scattered elastically off a hadron or the hadron decays weakly
into another hadron and an electron-antineutrino pair provide a
lot of information about the composition of hadrons in terms of
their constituents. The electric and weak coupling constants are
small enough such that leading-order perturbation theory suffices
to get meaningful results for the scattering and decay
probabilities. In leading-order the invariant scattering or decay
amplitudes become just contractions of a leptonic with a hadronic
current (times a $\gamma$, or $W$ propagator). The most general
covariant decomposition of the hadronic current leads to the
introduction of form factors. These are Lorentz-invariant
functions of the 4-momentum transferred to the hadron. They contain all the information about the substructure of the hadron, i.e. the deviation from a point-like
hadron, and can be directly extracted from (polarized) scattering
or decay cross sections.

The theoretical challenge is now to relate the hadron current to
the currents of its constituents. Under Poincar\'e transformations
a hadronic current operator  should transform
covariantly~\cite{Keister:1991sb}. Since the binding interaction shows up in some of the Poincar\'e generators, depending on the form of relativistic
dynamics to be used, it follows that $\hat{J}^\mu(x)$ must, in
general, also depend on the binding interaction and cannot be a
simple sum of the constituent currents. Further constraints for a
theoretical model of a hadron current come from current
conservation, i.e. $\partial_\mu \hat{J}^\mu(x) = 0$, and from the
requirement that the hadron charge should be the sum of the
constituent charges, independent on whether the binding
interaction is present or not.

We are primarily interested in calculating electroweak form
factors of strongly bound few-body systems within the framework of
relativistic quantum mechanics. A common procedure is to calculate
the bound-state wave function for a given binding force and use it
to {\em construct} a model for only the minimum number of current
components that is needed to fix the form factors uniquely. The
remaining current components are then determined by covariance and
current conservation. In usual front-form, e.g., it suffices to know the
$\hat{J}^+$ component if one calculates the current in the
$q^+=q^-=0$ frame~\cite{Keister:1991sb}. This kind of procedure,
however, has the drawback that the results for the form factors
will, in general, slightly depend on the chosen current components and
on the reference frame in which the construction of the current is
done~\cite{Carbonell:1998rj}. Our strategy is rather to {\em
derive} a full 4-vector current that is compatible with the
binding forces and valid in any reference frame within a
Poincar\'e invariant quantum mechanical setting. As it turns out,
such a current exhibits some unphysical features which, however,
can be split off in a unique way leaving a 4-vector current with
all the desired properties. Surprisingly, the outcome of our
approach resembles very much the results obtained within the
covariant light-front formalism that was suggested in
Ref.~\cite{Carbonell:1998rj}. This is the more remarkable since we
use the point form of relativistic quantum mechanics. In this form
all components of the 4-momentum become interaction dependent,
whereas the Lorentz generators stay free of interactions. This
guarantees simple boost and covariance properties of wave
functions and physical observables, respectively.

\section{Relativistic multichannel formalism and hadron currents}
\label{sec:1}
Our derivation of electroweak form factors starts with the
physical  processes in which the form factors are measured, i.e.
elastic electron-hadron scattering and the weak decay of hadrons.
We describe these reactions by means of a coupled-channel
framework in which the dynamics of the intermediate gauge bosons
-- either a photon or a W-boson -- is fully taken into account.
Poincar\'e invariance is ensured by employing the Bakamjian-Thomas
construction~\cite{Bakamjian:1953kh}. In its point-form version
the (interacting) 4-momentum operator $\hat{P}^\mu$ is factorized
into an interacting mass operator and a free 4-velocity operator
$\hat P^\mu =\hat M \hat V^\mu_{\mathrm{free}}\, .$ It is
therefore only necessary to study an eigenvalue problem for the
mass operator.

We will exemplify our approach through elastic electron-meson
scattering with the meson being a spin-0 confined quark-antiquark
system. In this case  a mass eigenstate $\hat{M} |\psi \rangle = m
|\psi \rangle$ is written as a direct sum of a
quark-antiquark-electron component $|\psi_{q\bar{q} e} \rangle$
and a quark-antiquark-electron-photon component $|\psi_{q \bar{q}
e\gamma} \rangle$. The mass eigenvalue equation to be solved has
then the form
\begin{eqnarray}\label{eigenvalue:equation}
 \left(\begin{array}{ll} \hat M_{q\bar{q} e} & \hat K_{\mathrm{em}}
 \\ \hat K_{\mathrm{em}}^\dagger &
 \hat M_{q\bar{q} e\gamma}\end{array}\right)
 \left(\begin{array}{l}  |\psi_{q\bar{q} e}\rangle
 \\   |\psi_{q\bar{q} e\gamma}\rangle
 \end{array}\right)   =
 m \left(\begin{array}{l} |\psi_{q\bar{q} e}\rangle
 \\ |\psi_{q\bar{q} e\gamma}\rangle
 \end{array}\right)\, ,
\end{eqnarray}
where $M_{q\bar{q} e}$ and $M_{q\bar{q} e\gamma}$ consist of a
kinetic term and an instantaneous confining potential between
quark and antiquark, and $\hat K^{(\dag)}_{\mathrm{em}}$ is a vertex operator which
accounts for the absorption (emission) of a photon by the
electron or (anti)quark. It is derived from the interaction
Lagrangean density of QED~\cite{Klink:2000pp}.

The current matrix elements that are necessary for the calculation
of the electromagnetic meson form factors can be extracted from
the invariant 1-photon-exchange amplitude. This is essentially
given by the on-shell matrix elements of the optical potential
$\hat{V}_{\mathrm{opt}}(m):=\hat K_{\mathrm{em}} (\hat M_{q\bar{q}
e\gamma}-m)^{-1}\hat K^\dagger_{\mathrm{em}}$. These matrix elements exhibit
the expected structure:
\begin{eqnarray}\label{eq:contract} \mathcal{M}_{1\gamma}
(\vec{k}_e^\prime, \mu_e^\prime;\vec{k}_e, \mu_e) &\propto&
\langle V^\prime; \vec{k}_e^\prime, \mu_e^\prime; \vec{k}_M^\prime
\vert \hat{V}_{\mathrm{opt}}(m) \vert V; \vec{k}_e, \mu_e;
\vec{k}_M\rangle_{\mathrm{on-shell}}\nonumber\\ & \propto & V^0
\delta^3(\vec{V}-\vec{V}^\prime)\frac{j_{{\mathrm{em}}\,\mu}
(\vec{k}_e^\prime, \mu_e^\prime; \vec{k}_e, \mu_e)
J^\mu_{\mathrm{em}}(\vec{k}_M^\prime;\vec{k}_M)}{(k_e^\prime-k_e)^2}\,
\end{eqnarray}
such that the meson current
$J^\mu_{\mathrm{em}}(\vec{k}_M^\prime;\vec{k}_M)$ can be easily
identified. $\vert V^{(\prime)}; \vec{k}^{(\prime)}_e, \mu^{(\prime)}_e;
\vec{k}^{(\prime)}_M\rangle$ are, so called, \lq\lq velocity
states\rq\rq\ that specify the state of a system by its overall
velocity $V^{(\prime)}$, the center-of-mass momenta $\vec{k}_i^{(\prime)}$ and
the canonical spins $\mu_i^{(\prime)}$ of its
components~\cite{Klink:1998zz}. In our case $\vec{k}^{(\prime)}_M$
is the momentum of the confined $q$-$\bar{q}$ subsystem with the
quantum numbers of the meson. \lq\lq On-shell" means that
$m=k_e^0+k_M^0=k_e^{\prime \,0}+k_M^{\prime\, 0}$ and
$k_e^0=k_e^{\prime \,0}$, $k_M^0=k_M^{\prime \,0}$. A detailed
derivation of Eq.~(\ref{eq:contract}) and the explicit expression
for the meson current
$J^\mu_{\mathrm{em}}(\vec{k}_M^\prime;\vec{k}_M)$ in terms of the
constituent currents and the bound-state wave functions are
given in Refs.~\cite{Biernat:2009my,Biernat:2011,Gomez:2011}.

It is quite obvious, how this formalism can be generalized to
obtain an expression for the weak meson transition current
$J^\mu_{\mathrm{wk}}(\vec{k}_{M^\prime}^\prime;\vec{k}_M)$ that
enters the semileptonic $M\rightarrow M^\prime e^- \bar{\nu}_e$
decay. The leading-order invariant transition amplitude ${\mathcal
M}_{1W}$ in this case can be derived from a 4-channel problem. In
addition to the incoming $q\bar{q}$ channel and the outgoing
$q^\prime\bar{q}e\bar{\nu}_e$ channel one needs a
$q^\prime\bar{q}W$ and a $q\bar{q}We\bar{\nu}_e$ channel to
account for the intermediate states in which the $W$-boson is in
flight. Here we assume that the flavor
change due to the coupling
of the $W$-boson happens for the quark. ${\mathcal M}_{1W}$ is
again given by on-shell matrix elements
($m=m_M=k_M^0=k_{M^\prime}^{0\prime}+k_e^{0\prime}+
k_{\bar{\nu}_e}^{0\prime}$) of the optical transition potential
$\hat{V}_{\mathrm{opt}}^{M\rightarrow M^\prime}\!\!(m):=\hat
K_{\mathrm{wk}} (\hat M_{q^\prime\bar{q}W}-m)^{-1}\hat
K^\dagger_{\mathrm{wk}}+\hat K_{\mathrm{wk}} (\hat
M_{q\bar{q}We\bar{\nu}_e}-m)^{-1}\hat K^\dagger_{\mathrm{wk}}$.
The vertex operators $K^{(\dagger)}_{\mathrm{wk}}$ for the
absorption (emission) of the $W$ by the quarks and leptons are
derived from the interaction Lagrangean density of
QFD~\cite{Klink:2000pp}. The mass operators $M_{q^\prime\bar{q}W}$
and $M_{q\bar{q}We\bar{\nu}_e}$ contain again an instantaneous
confining potential between quark and antiquark. The invariant
transition amplitude ${\mathcal M}_{1W}$ has the same structure as
the 1-photon-exchange amplitude ${\mathcal M}_{1\gamma}$ (cf.
Eq.~(\ref{eq:contract})) with the electromagnetic currents
replaced by the weak currents and the photon propagator
$(k_e^\prime-k_e)^{-2}$ replaced by the (covariant) $W$-propagator
$((k_e^\prime+k^\prime_{\bar{\nu}_e})^2-m_W^2)^{-1}$. A detailed
derivation and the explicit expression for the weak meson
transition current
$J^\mu_{\mathrm{wk}}(\vec{k}_{M^\prime}^\prime;\vec{k}_M)$ in
terms of the quark current and the bound-state wave functions can
be found in Ref.~\cite{Gomez:2011}.

\section{Electromagnetic current and form factors}
\label{sec:2}
\begin{figure}[t!]
  \includegraphics[width=0.48\textwidth]{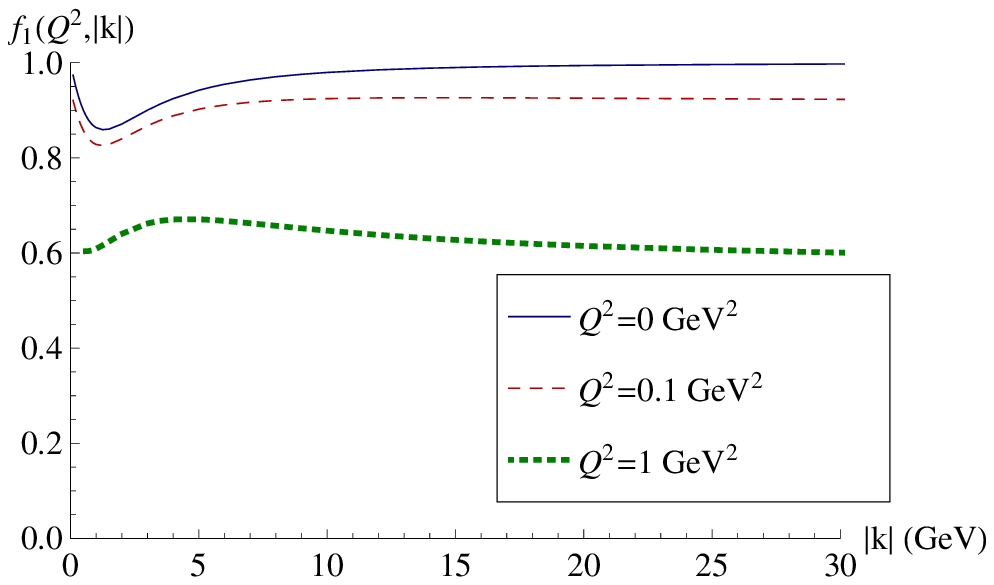}
  \hfill\includegraphics[width=0.48\textwidth]{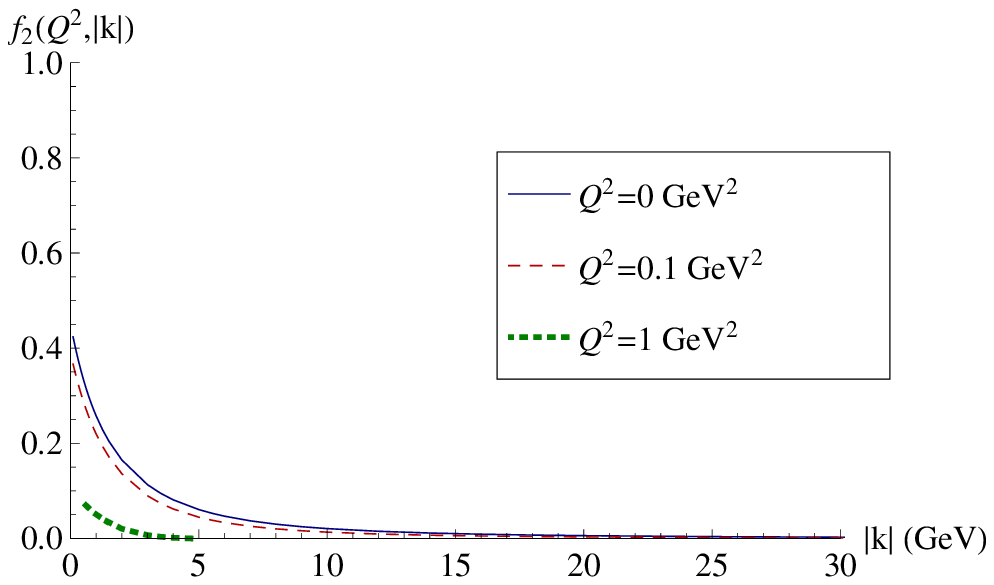}\vspace{-0.2cm}
\caption{Dependence of the electric $D$-meson form factor $f_1$ and
the spurious $D$-meson form factor $f_2$ (cf.
Eq.(\ref{physical:spurious})) on the meson center-of-mass momentum
$k=|\vec k_M|$ for different values of the momentum transfer
$Q^2$.}
\label{fig:sdep}\vspace{-0.3cm}
\end{figure}

As a next step we will analyze the properties of the
electromagnetic   meson current
$J^\mu_{\mathrm{em}}(\vec{k}_{M}^\prime;\vec{k}_M)$ that
follows from Eq.~(\ref{eq:contract}). Since we are using velocity
states in which $\vec k_M$ and $\vec k'_M$ are always defined in
the center-of-mass of the electron-meson system,
$J^\mu_{\mathrm{em}}(\vec k'_M;\vec k_M)$ does not transform like
a 4-vector under a Lorentz transformation $\Lambda$, but it rather
transforms by the Wigner rotation $R_W(V,\Lambda)$. A current with
the correct transformation properties is obtained by going back to
the physical meson momenta $p_M^{(\prime)}=B_c(V)k_M^{(\prime)}$,
i.e. by boosting the center-of-mass momenta with the overall
velocity $V$ of the electron-meson system:
\begin{equation}\label{eq:jcov}
J^\mu_{\mathrm{em}}(\vec p'_M;\vec p_M):=[B_c(V)]^\mu_\nu
J^\nu_{\mathrm{em}}(\vec k_M';\vec k_M)\, .
\end{equation}
$B_c(V)$ is the boost matrix of a rotationless boost. If $M$  is a
pseudoscalar meson it can be shown that $J^\mu_{\mathrm{em}}(\vec
p'_M;\vec p_M)$ is conserved, i.e. $(p'_M-p_M)_\mu
J^\mu_{\mathrm{em}}(\vec p'_M;\vec
p_M)=0$~\cite{Biernat:2009my,Biernat:2011}. Since internal momenta
are integrated over, the most general covariant decomposition of
$J^\mu_{\mathrm{em}}$ thus takes on the form
\begin{equation}\label{physical:spurious}
 J^\mu_{\mathrm{em}}(\vec p'_M;\vec p_M)=
 (p_M+p'_M)^\mu f_1(Q^2,s)+(p_e+p'_e)^\mu f_2(Q^2,s)\, ,
\end{equation}
with $Q^2=-(p'_M-p_M)^2$ and $s=(p_M+p_e)^2$. With the
Bakamjian-Thomas construction we have a nice Poincar\'e invariant
treatment of the electron-meson system, but the price we pay is a
dependence of the meson current on the electron momenta which
should not be there. It is the consequence of wrong cluster
properties, a well known drawback of the Bakamjian-Thomas
construction~\cite{Keister:1991sb}. As numerical studies
reveal, however, the dependence on the electron momenta becomes negligible
if the invariant mass $\sqrt{s}$ of the electron-meson system is
taken large enough~\cite{Biernat:2009my,Biernat:2011}. This is
demonstrated in Fig.~\ref{fig:sdep} for the case of a $D^+$ meson
and a simple Gaussian $\psi(\kappa)\propto \exp(-\kappa^2/(2
a^2))$, $a=0.55$~GeV taken for the bound-state wave function (the
quark masses are $m_c=1.6$~GeV, $m_{u,d}=0.25$~GeV).

In order to get rid of the spurious $k_e^{(\prime)}$ dependencies
it is thus tempting to take the limit $s\rightarrow\infty$. In
this limit the electromagnetic current of a pseudoscalar meson
acquires indeed its usual form
$J^\mu_{\mathrm{em}}(\vec{k}_M^\prime;\vec{k}_M)=({k}_M^\prime+{k}_M)^\mu
F(Q^2)$ and the analytical expression for the electric form factor
becomes rather simple. For equal quark and antiquark masses and
pure s-wave pseudoscalar mesons it is given
by~\cite{Biernat:2009my,Biernat:2011}
\begin{equation} \label{eq:FFps}
F(Q^2)=\lim_{s\rightarrow \infty} f_1(Q^2,s)=\frac{1}{4\pi} \int
\mathrm{d}^3\tilde{k}^\prime\,
\sqrt{\frac{\tilde{k}^0}{\tilde{k}^{0\prime}}}\, \mathcal{S}\,
\psi^\ast(|\vec{\tilde{k}}^\prime|) \psi(|\vec{\tilde{k}}|)\, ,
\end{equation}
with the spin-rotation factor $\mathcal{S}$ being the trace of a
product of Wigner $D$ functions. Primed and unprimed momenta are
related by appropriate rotationless boosts,
$\tilde{k}=\lim_{s\rightarrow\infty}
B_c^{-1}(v_{q\bar{q}})[B_c(v^\prime_{q\bar{q}})
\tilde{k^\prime}+k_M-k^\prime_M]$. Remarkably, by a simple change
of integration variables the form factor expression in
Eq.~(\ref{eq:FFps}) goes over into the standard front-form result
for a spectator current in the $q^+=0$
frame~\cite{Biernat:2009my,Biernat:2011} with $\mathcal{S}$
becoming the Melosh-rotation factor.
\begin{figure}[t!]
  \includegraphics[width=0.52\textwidth]{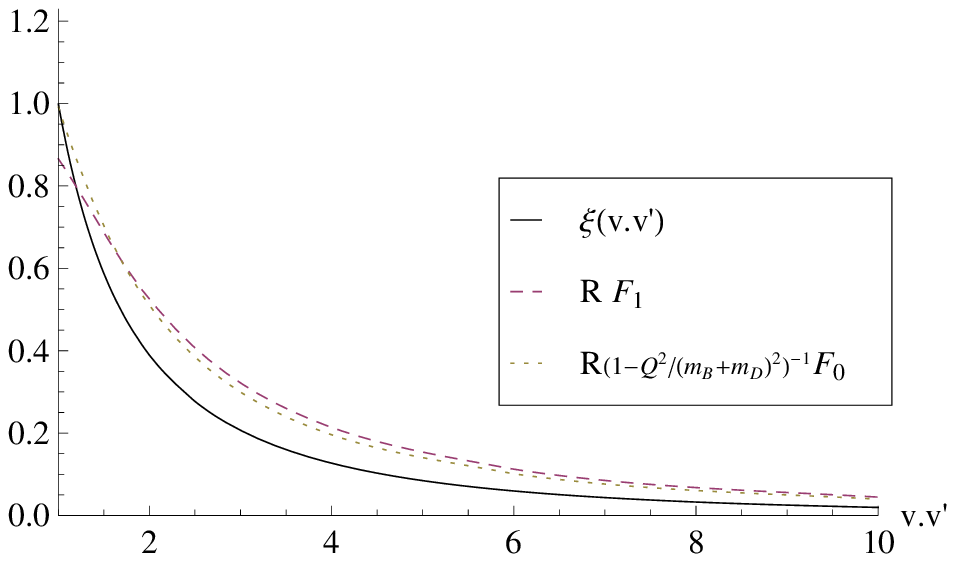}
  \includegraphics[width=0.52\textwidth]{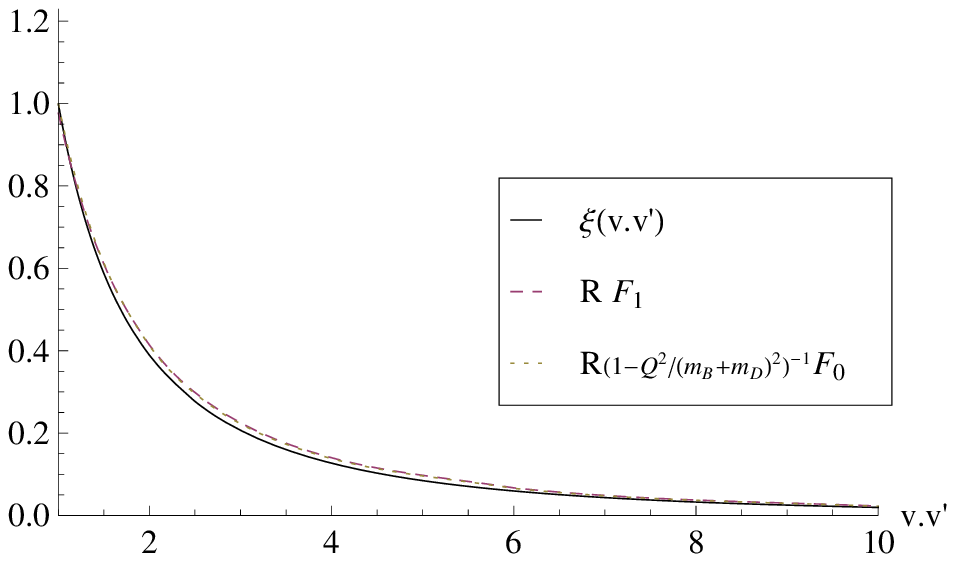}\vspace{-0.2cm}
\caption{Weak $B^- \rightarrow D^0 e^- \bar{\nu}_e$ decay form
factors (multiplied with appropriate kinematical factors) for
finite heavy-quark masses in comparison with the Isgur-Wise
function. Results for the physical $b$- and $c$-quark masses are
shown in the left figure. In the right figure these masses are
multiplied by a factor 10.}
\label{fig:2}\vspace{-0.3cm}       
\end{figure}

The generalization to pseudoscalar mesons with unequal-mass
constituents is straightforward~\cite{Gomez:2011}. What is
interesting in this connection is the heavy-quark limit (HQL) in
which the mass of the heavy constituent (say the quark) and hence
also of the meson goes to infinity. This limit has to be taken in
such a way that $v_M\cdot v^\prime_M=p_M\cdot p^\prime_M / m_M^2$
stays constant and $m_q=m_M$. In this limit the spurious
contributions are also seen to vanish and one finds
that~\cite{Gomez:2011}
\begin{equation}
J^\mu_{\mathrm{em}}(\vec p'_M;\vec
p_M)\stackrel{HQL}{\longrightarrow}\,  m_M \, (v_M+v^\prime_M)\,
\xi(v_M\cdot v^\prime_M)
\end{equation}
with the Isgur-Wise function~\cite{Isgur:1989vq}
\begin{eqnarray}\label{eq:xiem}
\xi(v_M\cdot v_M^\prime)= \frac{1}{4\pi}\int \mathrm{d}^3 \tilde{
k}^\prime_{\bar{q}}\, \sqrt{\frac{\tilde{
 k}_{\bar q}}{\tilde{
k}^{0\prime}_{\bar q}}}\, \sqrt{\frac{2}{1+v_M\cdot v'_M}}
\,\mathcal{W}\,
\psi^\ast(|\vec{\tilde k}'_{\bar
q}|)\, \psi(|\vec{\tilde k}_{\bar q}|)\, .
\end{eqnarray}
The Wigner rotation factor $\mathcal{W}$ is a function of
$\tilde{k}_{\bar{q}}^\prime$ and $(v_M\cdot v^\prime_M)$. Fig.~2 shows
the result for the Isgur-Wise function obtained with the same
(light) quark mass and the same Gaussian bound-state wave function
as in Fig.~1. This input has also been used in a front-form
calculation of the Isgur-Wise function~\cite{Cheng:1996if} and we
find indeed numerical agreement with the result of Cheng et al.

\section{Weak current and form factors}
\vspace{-0.2cm} Let us next turn to the analysis of the weak
current $J^\mu_{\mathrm{wk}}$ that can be extracted from the
semileptonic decay amplitude $\mathcal{M}_{1W}$. We will again
consider a pseudoscalar to pseudoscalar transition. As in the
electromagnetic case (cf. Eq.~(\ref{eq:jcov})) we have to go back
to the physical particle momenta to obtain a current that
transforms like a 4-vector. We have to note, however, that the
momentum transferred to the meson is now timelike, whereas it is
spacelike in electron-meson scattering. Interestingly, wrong
cluster properties of the Bakamjian-Thomas construction do not
entail unphysical properties of the decay current. Therefore its
covariant decomposition takes on the usual
form~\cite{Neubert:1993mb}:
\begin{equation}
J^\mu_{\mathrm{wk}}(\vec p'_{M^\prime};\vec p_M)=
 \left((p_M+p'_{M^\prime})^\mu-\frac{m_M^2-m_{M^\prime}^2}{q^2} \right) F_1(q^2)+
 \frac{m_M^2-m_{M^\prime}^2}{q^2}q^\mu F_0(q^2)\, ,
\end{equation}
with $q=(p_M-p'_{M^\prime})$. If heavy-quark (flavor) symmetry holds the
heavy-quark limit of $F_0$ and $F_1$ (multiplied with appropriate
kinematical factors) should give the same (universal) Isgur-Wise
function as the heavy-quark limit of the electromagnetic form
factor ($q^2$ again replaced by ($v^\prime_{M^\prime}\cdot v_M$)).
This is indeed the case, which proves that our procedure to calculate
currents and form factors respects heavy-quark symmetry. But we
can also calculate the form factors for finite (physical) quark
masses to estimate how far nature is away from the heavy quark
limit. The result for the $B^- \rightarrow D^0 e^- \bar{\nu}_e$
decay is plotted in Fig.~\ref{fig:2}. For physical heavy-quark
masses the deviation is sizable. Approximate restauration of
heavy-quark symmetry is, however, observed for masses that are
about 10 times larger.

\section{Concluding remarks}
%
We have seen for spin-0 2-particle bound states and instantaneous
binding forces that our point-form approach provides results for
electromagnetic form factors that agree with front form
calculations in the $q^+=0$ frame. We agree also in the spin-1
case with the outcome of the covariant front-form approach of
Carbonell et al.~\cite{Carbonell:1998rj}. This is discussed
elsewhere~\cite{Biernat:2011}. We have further calculated weak
decay form factors for heavy-light systems. Our formalism is seen
to give the right heavy-quark limit with the Isgur-Wise function
being again in agreement with the front form result. This has also
been checked for pseudoscalar to vector
transitions~\cite{Gomez:2011}, which verifies heavy-quark spin
symmetry. What remains to be seen is, whether the agreement with
form-factor calculations in front form will continue to hold for
binding forces caused by dynamical particle exchanges.


\begin{acknowledgements}
M. G\'omez-Rocha acknowledges the support of the \lq\lq Fonds zur
F\"orderung der wissenschaftlichen Forschung in \"Osterreich" (FWF
DK W1203-N16).
\end{acknowledgements}


\end{document}